\documentclass[12pt,a4paper,reqno,oneside]{amsart}%

\usepackage[a4paper,DIV12]{typearea}
\usepackage{amssymb}
\usepackage{pstricks}
\usepackage{xspace}		% \xspace - automatically add space after macros
\usepackage{longtable}	        % long tables
\usepackage{booktabs}		% \toprule, \midrule, \bottomrule
\usepackage{mathtools}          % \math[rl]lap, \shortintertext, \coloneqq ...
\mathtoolsset{mathic}           % show only referenced tags, italic corr. in math mode \( \)
\usepackage{enumitem}		% formatting of itemized lists, description lists, and enumerated lists

\usepackage[%draft, % do not do any hyper linking,
            dvips, colorlinks=true, linkcolor=black, urlcolor=black, citecolor=black,
            pdfpagelabels, bookmarks, bookmarksnumbered, naturalnames=true,
			pdfkeywords={on-line, partial order, chain partition},
            bookmarksopen, linktocpage,
			pdftitle={On-line Chain Partitions of Up-growing Semi-orders}, 
                        pdfdisplaydoctitle,
           ]{hyperref}

\makeatletter
\def\revddots{\mathinner{\mkern1mu\raise\p@
\vbox{\kern7\p@\hbox{.}}\mkern2mu
\raise4\p@\hbox{.}\mkern2mu\raise7\p@\hbox{.}\mkern1mu}}
\makeatother

\newcounter{simplificationcnt}

\newenvironment{enumeratei}{\begin{enumerate}[label=\textup{(\roman*)}, noitemsep, topsep=1.5mm, labelindent=.8em, leftmargin=*, widest=.]}{\end{enumerate}}
\newenvironment{enumeratenum}{\begin{enumerate}[label=\textup{(\arabic*)}, noitemsep, topsep=1.5mm, labelindent=.8em, leftmargin=*, widest=.]}{\end{enumerate}}

\newtheorem{theorem}{Theorem}[section]
\newtheorem{proposition}[theorem]{Proposition}

\newtheorem{lemma}[theorem]{Lemma}
\newtheorem{fact}[theorem]{Fact}

\theoremstyle{definition}

\renewcommand{\phi}{\varphi}

\renewcommand{\leq}{\leqslant}
\renewcommand{\geq}{\geqslant}

\newcommand{\phaseskip}{\smallskip}
\newcommand{\abs}[1]{\left\vert#1\right\vert}

\DeclareMathSymbol{\oUpset}{\mathclose}{symbols}{34}
\DeclareMathSymbol{\cUpset}{\mathclose}{symbols}{42}
\DeclareMathSymbol{\oDownset}{\mathclose}{symbols}{35}
\DeclareMathSymbol{\cDownset}{\mathclose}{symbols}{43}
\newcommand{\N}{\ensuremath{\mathbb{N}}\xspace}  
\newcommand{\set}[1]{\left\{#1\right\}}

\newcommand{\itemref}[1]{\ref{#1}}

\newcommand{\ThreePlusOne}{\ensuremath{\mathbf{(3 + 1)}}\xspace}
\newcommand{\TwoPlusTwo}{\ensuremath{\mathbf{(2 + 2)}}\xspace}
\newcommand{\PP}{\ensuremath{\mathbf{P}}\xspace}

\newcommand{\OO}{\ensuremath{\mathcal{O}}\xspace}
\newcommand{\CC}{\ensuremath{\mathcal{C}}\xspace}

\newcommand{\GAMMA}{\ensuremath{\Gamma}\xspace}

\DeclareMathOperator{\fHeight}{height}
\DeclareMathOperator{\fWidth}{width}
\DeclareMathOperator{\fTop}{top}
\DeclareMathOperator{\fBot}{bottom}

\newcommand{\Case}[1]{\smallskip\par\noindent\textbf{Case #1.\ }}

\newcommand{\phase}[1]{\smallskip\par\noindent\textbf{Phase}~$#1$.~}

\DeclareMathOperator{\fVal}{val}
\DeclareMathOperator{\fInc}{Inc}
\DeclareMathOperator{\alg}{alg}

\newcommand{\ALG}{\textnormal{ALG}\xspace}

\def\ITEMMACRO #1 ??? #2 ???{\par\vskip4pt\noindent%
\hangindent=#2em\setbox0\hbox{#1\kern4pt}%
\ifdim\wd0<\hangindent\setbox0\hbox to\hangindent{\hss#1\kern7pt}\fi%
\box0\ignorespaces}

\def\Item(#1){\ITEMMACRO {\rm (#1)} ??? 1.8 ???}
\def\FreeItem#1{\ITEMMACRO {#1} ??? 1.8 ???}

\def\BrackItem[#1]{\ITEMMACRO [#1] ??? 1.8 ???}

\def\A#1,#2.{{}${\sf above}$_{#1,#2}}
\def\L#1,#2.{{}${\sf left}$_{#1,#2}}
\def\R#1,#2.{{}${\sf right}$_{#1,#2}}

\def\ld#1,#2,#3.{#1_{#2},\ldots,#1_{#3}}

\def\rih#1{\noindent{\bf #1}}

\def\Case#1.{\rih{Case #1.}}
\def\Phase#1.{\rih{Phase #1.}}
\def\Fact#1.{\rih{Fact #1.}}

\def\NN{\hbox{\sf I\kern-1ptN}}

\begin{document}

% *************************************************************
%                   TITLE PAGE
% *************************************************************

\title{On-line Chain Partitions of Up-growing Semi-orders}%
\author[S. Felsner]{Stefan Felsner}%
\address{AG Diskrete Mathematik\\
   Institut f\"{u}r Mathematik\\
   Technische Universit\"{a}t Berlin\hfil\break\indent
   Strasse des 17. Juni 136, 10623 Berlin, Germany}
\email{felsner@math.tu-berlin.de}%
\urladdr{\url{http://www.math.tu-berlin.de/~felsner/}}%
\author[K. Kloch]{Kamil Kloch}
\address{Embedded Systems Lab, University of Passau, 
        Innstrasse 43, 94032 Passau, \mbox{Germany}}

\email{kamil.kloch@uni-passau.de}%
\urladdr{http://www.esl.fim.uni-passau.de/\~{}kloch/}%

\author[G. Matecki]{Grzegorz Matecki}
\author[P. Micek]{Piotr Micek}
\address{
  Algorithmics Research Group\\
  Faculty of Mathematics and
  Computer Science\hfil\break\indent
  Jagiellonian University\\
  Lojasiewicza 6, 30-348 Krak\'{o}w, Poland}%
\email{Grzegorz.Matecki@tcs.uj.edu.pl}%
\urladdr{\url{http://tcs.uj.edu.pl/Matecki}}%
\email{Piotr.Micek@tcs.uj.edu.pl}%
\urladdr{\url{http://tcs.uj.edu.pl/Micek}}%
\date{\today}%

\begin{abstract}
On-line chain partition is a two-player game between Spoiler and
Algorithm. Spoiler presents a partially ordered
set, point by point. Algorithm assigns incoming points (immediately and irrevocably) to the
chains which constitute a chain partition of the order. The value of the game
for orders of width $w$ is a minimum number $\fVal(w)$ such that Algorithm has
a strategy using at most $\fVal(w)$ chains on orders of width at most $w$.  We
analyze the chain partition game for up-growing semi-orders. Surprisingly, the
golden ratio comes into play and the value of the game is $\lfloor\frac{1+\sqrt{5}}{2}\; w \rfloor$.
\end{abstract}

\maketitle

%&%&%&%&%&%&%&%&%&%&%&%&%&%&%&%&%&%&%&%&%&%&%&%&%&%&%&%&%&%&%&%&%&%&%&%&%&%&%&%

% *************************************************************
\section{Introduction}
% *************************************************************
On-line chain partitions of an order can be described as a two-person
game between Algorithm and Spoiler. The game is played in
rounds. Spoiler presents an on-line order, one point at a
time. Algorithm responds by making an irrevocable assignment of the
new point to one of the chains of the chain partition. The performance
of Algorithm's strategy is measured by comparing the number of chains
used with the number of chains of an optimal chain partition.
By Dilworth's Theorem the size of an optimal chain partition
equals the width of the order. The value of
the game for orders of width $w$, denoted by $\fVal(w)$, is the
least integer $n$ for which some Algorithm has a strategy using
at most $n$ chains for every on-line order of width $w$. 
Alternatively, it is the largest integer $n$ for which Spoiler
has a strategy that forces any Algorithm to use $n$ chains on
order of width $w$.

The study of chain partition games goes back to the early 80's when
Kierstead~\cite{Kier81} (upper bound) and Szemer\'{e}di (lower bound published
in \cite{Kier86}) proved the estimates for on-line orders of width $w$:
$\binom{w+1}{2} \leqslant \fVal(w) \leqslant \frac{5^w-1}{4}$. 
It took almost 30 years until these bounds had been slightly improved.
The story can be found in the survey~\cite{FKMM-survey}.

The study of on-line chain partition on restricted classes of orders
began in 1981 when Kierstead and Trotter \cite{KierTro81} proved the following result: when Spoiler is restricted to presenting interval orders of width $w$, the value of the game is $3w-2$.
Among other classes of orders that have been studied thereafter are
$(\mathbf{k+k})$-free orders and semi-orders. Again we refer 
to~\cite{FKMM-survey} for details. 

Up-growing on-line orders have been introduced by Felsner~\cite{Fel97}. 
In this variant Spoiler's power is restricted by the
condition that the new element has to be a maximal element of the
order presented so far.  Felsner~\cite{Fel97} showed that the value of
the chain partition game on up-growing orders is $\binom{w+1}{2}$. The
case of up-growing interval orders was resolved by Baier,
Bosek and Micek~\cite{BBM07}. The value of the 
game in this variant is $2w-1$. 

This paper resolves the on-line chain partition problem for up-growing semi-orders. An order $\PP$ is called a semi-order
if it has a unit interval representation, i.e., there exists a mapping $I$ of
points of the order into unit length intervals on a real line so that $x < y$
in $\PP$ iff interval $I(x)$ is entirely to the left of $I(y)$. Alternatively
semi-orders are characterized as the \TwoPlusTwo and \ThreePlusOne-free
orders (see Fig.~\ref{fig_2+2_1+3}).

Considering on-line chain partitions of semi-orders note that the general (not
up-growing) case is easy to analyze: First, observe that the number of chains
used by Algorithm can be bounded by $2w-1$.  Let $x$ be the new point and
consider the set $\fInc(x)$ of points incomparable with $x$.  Clearly, the
only chains forbidden for $x$ are those used in $\fInc(x)$. Now
$\fWidth(\fInc(x)) \leqslant w-1$ since the width of the whole order does not
exceed $w$. Moreover, $\fHeight(\fInc(x)) \leqslant 2$ as the presented order
is \ThreePlusOne-free. Therefore, $|\fInc(x)| \leqslant 2(w-1) = 2w-2$, 
proving that $x$ can be assigned to at least one of
$2w-1$ legal chains.

\begin{figure}[!ht]
    \begin{center} \ifx\JPicScale\undefined\def\JPicScale{1}\fi
\psset{unit=\JPicScale mm}
\psset{linewidth=0.3,dotsep=1,hatchwidth=0.3,hatchsep=1.5,shadowsize=1,dimen=middle}
\psset{dotsize=0.7 2.5,dotscale=1 1,fillcolor=black}
\psset{arrowsize=1 2,arrowlength=1,arrowinset=0.25,tbarsize=0.7 5,bracketlength=0.15,rbracketlength=0.15}
\begin{pspicture}(0,0)(87.5,27.5)
\rput(2.5,10){$a_1$}
\rput(5,12.5){$a_2$}
\rput(7.5,15){$a_3$}
\psline[linewidth=0.4,arrowsize=0 2,arrowlength=0,arrowinset=-0.25,tbarsize=0 5,bracketlength=0,rbracketlength=0]{|-|}(5,10)(23.75,10)
\psline[linewidth=0.4,arrowsize=0 2,arrowlength=0,arrowinset=-0.25,tbarsize=0 5,bracketlength=0,rbracketlength=0]{|-|}(7.5,12.5)(26.25,12.5)
\rput(28.75,25){$\revddots$}
\psline[linewidth=0.4,arrowsize=0 2,arrowlength=0,arrowinset=-0.25,tbarsize=0 5,bracketlength=0,rbracketlength=0]{|-|}(22.5,27.5)(41.25,27.5)
\psline[linewidth=0.4,arrowsize=0 2,arrowlength=0,arrowinset=-0.25,tbarsize=0 5,bracketlength=0,rbracketlength=0]{|-|}(17.5,22.5)(36.25,22.5)
\rput(21.25,17.5){$\revddots$}
\rput(12.5,20){$a_k$}
\psline[linewidth=0.4,arrowsize=0 2,arrowlength=0,arrowinset=-0.25,tbarsize=0 5,bracketlength=0,rbracketlength=0]{|-|}(15,20)(33.75,20)
\psline[linewidth=0.4,arrowsize=0 2,arrowlength=0,arrowinset=-0.25,tbarsize=0 5,bracketlength=0,rbracketlength=0]{|-|}(10,15)(28.75,15)
\rput(77.5,7.5){$b_1$}
\rput(80,10){$b_2$}
\rput(82.5,12.5){$b_3$}
\psline[linewidth=0.4,arrowsize=0 2,arrowlength=0,arrowinset=-0.25,tbarsize=0 5,bracketlength=0,rbracketlength=0]{|-|}(56.25,7.5)(75,7.5)
\psline[linewidth=0.4,arrowsize=0 2,arrowlength=0,arrowinset=-0.25,tbarsize=0 5,bracketlength=0,rbracketlength=0]{|-|}(58.75,10)(77.5,10)
\rput(72.5,15){$\revddots$}
\rput(87.5,17.5){$b_k$}
\psline[linewidth=0.4,arrowsize=0 2,arrowlength=0,arrowinset=-0.25,tbarsize=0 5,bracketlength=0,rbracketlength=0]{|-|}(66.25,17.5)(85,17.5)
\psline[linewidth=0.4,arrowsize=0 2,arrowlength=0,arrowinset=-0.25,tbarsize=0 5,bracketlength=0,rbracketlength=0]{|-|}(61.25,12.5)(80,12.5)
\rput(60,2.5){$\revddots$}
\psline[linewidth=0.4,arrowsize=0 2,arrowlength=0,arrowinset=-0.25,tbarsize=0 5,bracketlength=0,rbracketlength=0]{|-|}(53.75,5)(72.5,5)
\psline[linewidth=0.4,arrowsize=0 2,arrowlength=0,arrowinset=-0.25,tbarsize=0 5,bracketlength=0,rbracketlength=0]{|-|}(48.75,0)(67.5,0)
\psline[linewidth=0.4,arrowsize=0 2,arrowlength=0.5,arrowinset=0.1,tbarsize=0 5,bracketlength=0,rbracketlength=0]{|-|}(25,10)(57.5,10)
\psline[linewidth=0.4,arrowsize=0 2,arrowlength=0.5,arrowinset=0.1,tbarsize=0 5,bracketlength=0,rbracketlength=0]{|-|}(27.5,12.5)(60,12.5)
\psline[linewidth=0.4,arrowsize=0 2,arrowlength=0.5,arrowinset=0.1,tbarsize=0 5,bracketlength=0,rbracketlength=0]{|-|}(65,17.5)(32.5,17.5)
\rput(46.25,15){$\revddots$}
\rput(41.25,7.5){$x_1$}
\rput(42.5,12.5){$x_2$}
\rput(50,20){$x_{k-1}$}
\end{pspicture}
    \end{center}
    \caption{Strategy for Spoiler forcing Algorithm to use $2w-1$
chains.}
    \label{pic_2w-1}
\end{figure}
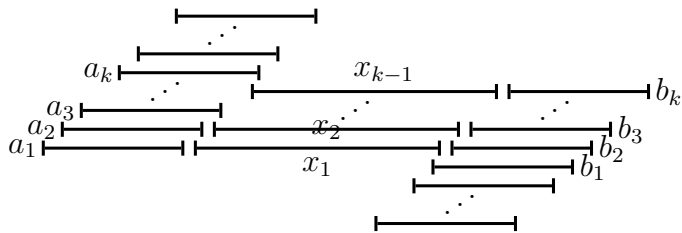

It turns out that there is no better strategy for Algorithm. In other
words, Spoiler may force Algorithm to use $2w-1$ chains on
semi-orders of width $w$. A strategy for Spoiler looks as follows:

\begin{enumeratenum}
	\item Present two antichains $A$ and $B$, both consisting of $w$
points in such a way that $A<B$, i.e., all points from $A$ are below
all points from $B$. If Algorithm uses $2w-1$ or more chains, the
construction is finished. Otherwise, suppose that $k$ chains ($2 \leqslant k \leqslant w$) contain elements from $A$ and~$B$, namely let $a_i\in A_i$, $b_i\in B_i$ for $1\leq i\leq k$ lie in the same chain.

	\item Present $k-1$ points $x_1, \ldots, x_{k-1}$ in such a way that $\set{a_1,\ldots,a_i}\leq x_i\leq\set{b_{i+1},\ldots,b_k}$ and $x_i$ is incomparable to all the rest (the interval representation of the whole order looks as in Fig.~\ref{pic_2w-1}).  It is easy to verify that in such setting
Algorithm is forced to use $2w-1$ chains.
\end{enumeratenum}

The contribution of this paper is the following theorem.
\begin{theorem}\label{thm-main}
The value of the on-line chain partition game for up-growing semi-orders
of width $w$ is $\lfloor \frac{1+\sqrt{5}}{2} \cdot w \rfloor$.
\end{theorem}

\section{Up-growing Semi-orders}\label{sec:semi-orders}
\subsection{Outline}
In this section we prove that the value of the on-line chain partition game
for up-growing semi-orders equals $\lfloor \phi \cdot w\rfloor$, where
$\phi=\frac{1+\sqrt{5}}{2}$ is the golden number.  First, in
Sect.~\ref{ssec:basics} we collect some facts about
semi-orders. Section \ref{ssec:lower-bound} describes a strategy for Spoiler
which forces Algorithm to use at least $\lfloor \phi \cdot w\rfloor$ chains on
a semi-order of width $w$. This sets the lower bound for the value of the
game. In Sect.~\ref{ssec:upper-bound} we propose a strategy for Algorithm
using at most $\lfloor\phi \cdot w\rfloor$ chains on semi-orders of width at
most $w$.

The presence of the golden number $\phi$ in the result of a chain partition
game may seem surprising. In fact, it is the Fibonacci sequence ($F_0=0$,
$F_1=1$ and $F_{i+2}=F_{i}+F_{i+1}$) which appears in the counting argument of
the upper bound and serves as a discrete counterpart of $\phi$.

% *************************************************************
\subsection{Basic Facts} \label{ssec:basics}%
% *************************************************************
For $x,y\in\PP$ by $x\parallel_{\PP} y$ we mean that $x$ and $y$ are incomparable in \PP. Let $x\oDownset_{\PP} = \{y \in P: y < x\}$, called a \emph{down set} of
$x$ in \PP, denote the set of predecessors of $x$ in \PP. Dually, let
$x\oUpset_{\PP} = \{y \in P: y > x\}$, called an \emph{up set} of $x$,
denote the set of successors of $x$ in \PP. If the order $\PP$ is
unambiguous from the context we also write $x\oUpset$ instead of
$x\oUpset_{\PP}$ and $x\oDownset$ instead of $x\oDownset_{\PP}$. By $X\oDownset$ we mean $\bigcup_{x\in X}x\oDownset$. 

The maximum and the minimum elements of a chain $\gamma$ are 
denoted respectively by $\fTop(\gamma)$ and $\fBot(\gamma)$.

An order \PP is called an \emph{interval order} if it has an interval
representation, i.e., there exists a mapping $I$ of points of the order into
intervals on a real line so that $x < y$ in \PP if{}f $\max I(x) < \min I(y)$.  Interval
orders have several nice characterizations, see e.g.~\cite{Moe89}.
In our context the following two will be used repeatedly:
\begin{enumeratenum}
\item $\PP=(P,\leqslant)$ is an interval order if{}f the set of
down sets (up sets) of elements of \PP\ is linearly ordered with respect to
inclusion,  i.e., for $p, q \in P$ either $p\oDownset \subseteq q\oDownset$
or $p\oDownset \supseteq q\oDownset$.
Note that this ordering of down sets 
corresponds to the order of left  endpoints in an interval
representation. 
\item
$\PP=(P,\leqslant)$ is an interval order if{}f \PP is a
\TwoPlusTwo-free order, i.e., \PP does not contain elements $a,b,c,d\in P$
such that: $a<b$, $c<d$, $a\parallel d$ and $c\parallel b$ (see
Fig.~\ref{fig_2+2_1+3}).
\end{enumeratenum}
\begin{figure}[!ht]
    \begin{center} \input{pictures/pic_22-13.tex}
    \end{center}
    \caption{\TwoPlusTwo\ and \ThreePlusOne\ orders.}
    \label{fig_2+2_1+3}
\end{figure}

An interval order $\PP$ is called a \emph{semi-order} if it has a unit
interval representation, i.e., all intervals are of the same length. Semi-orders
are also characterized in terms of forbidden structures: an interval order \PP
is a semi-order if{}f \PP is a \ThreePlusOne-free order, i.e., \PP does not
contain elements $e, f, g, h \in P$ such that $e < f < g$ and $h \parallel e,
f, g$ (see Fig.~\ref{fig_2+2_1+3}).

% *************************************************************
\subsection{The Lower Bound}\label{ssec:lower-bound}
% ************************************************************* 

Fix $w$ and consider
the system \eqref{equ-Ik} of $k$ linear inequalities
\begin{equation} \label{equ-Ik} \tag{$I_k$}
x_0 + x_1 + \ldots + x_{j-1} +2x_j-x_{j+1}\leqslant w,\ j=0,\ldots,k.
\end{equation}
From the following two propositions it immediately follows that there exists a strategy for Spoiler which forces Algorithm to use $\lfloor\phi\cdot w\rfloor$
chains on an up-growing semi-order of width $w$. This is the lower
bound needed for Theorem~\ref{thm-main}.

\begin{proposition}\label{prop:strategy-lower}
If $(x_0,x_1,\ldots,x_k,x_{k+1})$ is an integral solution of \eqref{equ-Ik}
with $x_0 \geq x_1\geq\ldots\geq x_k\geq x_{k+1}=0$ 
then there is a strategy for Spoiler to present an up-growing semi-order of width $w$ and force Algorithm to use at least $w+x_0$ chains.
\end{proposition}

\begin{proposition}\label{prop:solution_I_k}
For each $w$ there is an integer $k$ and an integral solution of
\eqref{equ-Ik} with $x_0=\lfloor(\phi-1)\cdot w\rfloor\geq
x_1\geq\ldots\geq x_k>x_{k+1}=0$. 
\end{proposition}

% *************************************************************
% \noindent{\sl Proof of Proposition~\ref{prop:strategy-lower}.}
\begin{proof}[Proof of Proposition~\ref{prop:strategy-lower}.]
Fix $w$, $k> 0$ and an integer solution $(x_0,\ldots,x_k)$ of \eqref{equ-Ik}
with $x_0 \geq x_1\geq\ldots\geq x_k\geq x_{k+1}=0$. The strategy for Spoiler induced
by $(x_0,\ldots,x_k)$ presents an up-growing semi-order $\PP=(P,\leqslant)$ of
width $w$. The height of \PP is at most $3$. The points of \PP are presented in bundles so that the actual
presentation sequence has the following structure
\[
P = (A,\,C_0,\,B_1,\,C_1,\,B_2,\,C_2,\,\ldots,\,C_{k},\,B_{k+1}). 
\]
The set $A$ is exactly the set of minimal elements of \PP. Points of height $2$ lie in $C_0\cup\bigcup_{i=1}^{k+1} B_i$ and all points from $\bigcup_{i=1}^k C_i$ are of height $3$.

Throughout the construction Spoiler maintains auxiliary sets $D_1, \ldots, D_k$.
Initially $D_i=\emptyset$ for every $i$. During the construction
the following invariant will be kept:
\begin{equation} \label{equ-invariant-D}
\textnormal{
$D_i\subseteq B_i$ and $D_i$ does not contain top of any chain used by Algorithm.}
\end{equation}

Now, we describe the phases of the construction. The proof that the construction
has all desired properties will follow thereafter. 
%An example of the entire game is shown in Fig.~\ref{figure:example-of-the-game}.
\phaseskip

Spoiler starts the construction by presenting an antichain $A$ of size $w$. Algorithm has
to use $w$ different chains.

% Next, Spoiler presents a set $B_0$ of
% size $x_0-x_1$ such that $B_0>A$ (i.e., $b>a$ for every $b\in B_0$
% and $a\in A$). Algorithm uses for $B_0$ chains from $A$ or/and
% new chains. Let $A_0\subseteq A$ be the set of points whose chains
% have been used in $B_0$. Clearly, $\abs{A_0}\leqslant x_0-x_1$. Now,
% Spoiler presents a set $B_1$ of size $x_0-x_1$ such that $A_0\subseteq
% B_1\oDownset$ and $\abs{B_1\oDownset}=x_0-x_1$. This is
% possible as $0\leq x_0-x_1\leqslant w=\abs{A}$, by \eqref{equ-Ik}.

\phaseskip

\phase{j\ (0\leqslant j\leqslant k)} In the $j$-th phase Spoiler
builds $x_{j}-x_{j+1}$ \emph{forcing paths}. The points constituting these paths will go into the set $C_{j}$.

The first point $q_0$ of a forcing path dominates $A\cup\bigcup_{i\leq j}  B_i$. Now, suppose that the first $i+1$ points of
the path have been presented and let $q_i$ be the last of these points. If
Algorithm assigned $q_i$ to a new chain or to a chain whose top is in $A$, then $q_i$ is the last point of the forcing path. Otherwise, $q_i$ was
assigned to some chain with a top $b\in B_s$. In this case Spoiler
updates $D_s\coloneqq D_s\cup\set{b}$ and then introduces
$q_{i+1}>A\cup B_1\cup\ldots\cup B_{s-1}\cup D_s$.

Note that the invariant \eqref{equ-invariant-D} is kept, i.e., $D_i$
does not contain any chain top. Algorithm has to assign $q_{i+1}$ to a
new chain or to a chain with top in $A\cup B_1\cup\ldots\cup
B_{s-1}$. This means that if $q_{i+1}$ is assigned to a chain with a
top from $B_{s'}$ then $s'<s$. Thus, consecutive points
$q_0,q_1,\ldots$ of a forcing path (excluding the last one) are
assigned to chains with tops from the $B_i$'s with decreasing indices.
This proves that the path is finite.

The intuition is that with each forcing path Spoiler forces Algorithm to produce a \emph{skip chain}, i.e., a chain of height 2 with its bottom in $A$ and its top in $C_j$ (avoiding all the $B_i$'s), or to use a brand new chain.

Assume that Spoiler constructed all the forcing paths and consider the set
$A_{j}\subseteq A$ of bottom points of skip chains with tops in $C_j$. Clearly,
$\abs{A_{j}}\leqslant x_{j}-x_{j+1}$. Now, Spoiler introduces a set
$B_{j+1}$ consisting of $x_{j}-x_{j+1}$ points such that $A_{j}\cup
B_{j}\oDownset\subseteq B_{j+1}\oDownset$ and
$\abs{B_{j+1}\oDownset}=x_0-x_{j+1}$ (put $B_0=\emptyset$). This means that if
$\abs{A_{j}}<x_j-x_{j+1}$ or $A_{j}\cap B_{j}\oDownset\neq\emptyset$
then Spoiler completes $B_{j+1}\oDownset$ with arbitrarily chosen
points from $A-(A_{j}\cup B_{j}\oDownset)$. This is possible as
$\abs{B_{j}\oDownset}+\abs{A_{j}}\leq(x_0-x_j)+(x_j-x_{j+1})=x_0-x_{j+1}\leqslant
w=\abs{A}$, by \eqref{equ-Ik}.

% \begin{figure}[!ht]
%     \begin{center} \input{pictures/pic-low-b.tex} \end{center}
%     \caption{Spoiler plays a strategy induced by $(8,\,3,\,1,\,0)$ for
%       $w=13$, i.e., presents a semi-order of width $13$ and forces
%       Algorithm to use $13+8=21$ chains. The first three pictures
%       illustrate the situation after Phases $0$, $1$ and $2$. Relation
%       $A < C$ is not visualized for the sake of clarity. The bottom
%       picture shows the unit-length interval representation of the
%       presented semi-order.} \label{figure:example-of-the-game}
% \end{figure} 
% 
% \medskip

To prove that the construction actually works and thus concludes the proof of
Proposition~\ref{prop:strategy-lower} we have to verify the
following three facts.
\begin{fact}\label{fact-P-is-semi-order} \PP is a semi-order.\end{fact}
\begin{fact}\label{fact-width-P-is-w} The width of \PP is $w$.\end{fact}
\begin{fact}\label{fact-number-of-chains-forced} Algorithm has to use at least $w+x_0$ chains to cover \PP.\end{fact}
\begin{proof}[Proof of Fact \ref{fact-P-is-semi-order}]
We proceed in two steps, first we show that \PP is an interval order
and then that it is \ThreePlusOne-free.

In order to prove that \PP is an interval order we show that the down sets of
points from \PP are linearly ordered with respect to
inclusion. Indeed,
\begin{enumeratei}
\item $A\oDownset=\emptyset$,
\item $B_1\oDownset\subseteq B_2\oDownset\subseteq\ldots\subseteq B_k\oDownset\subseteq A$,
\item $C_0\oDownset=A$,
\item if $c\in C_j$ is the starting point of a forcing path, then
  $c\oDownset=A\cup(B_1\cup\ldots\cup B_j)$,
\item if $c\in C_j$ is not a starting point,
  then there is an $s$ with $c\oDownset=A\cup(B_1\cup\ldots\cup B_{s-1}\cup D^t_{s})$.
\end{enumeratei}

\noindent The set $D^t_{s}$ from the last line refers to the respective
set $D_s$ at the moment when $c$ is introduced. Recalling that 
$D_s$ can only grow over time
and $D_s\subseteq B_s$ we can conclude that the down sets of 
elements of $c$ are linearly ordered with respect to
inclusion. Hence \PP is an interval order. 

To see that \PP is a semi-order
suppose that \PP contains a $\ThreePlusOne$-configuration
${d}\parallel{a,\,b,\,c}$ with $a<b<c$. Since \PP has height at most
$3$ and $A<\bigcup_{i=0}^{k} C_i$, the only option is that %$d\in B$ while 
$a\in A$, $b,d\in \bigcup_{i=1}^{k+1} B_i$ and $c\in \bigcup_{i=1}^{k} C_i$ ($C_0$ is excluded as it is incomparable to the $B_i$'s). Let $i,\,j$ be such that $b\in B_i$ and
$d\in B_j$. Then it is easy to see that $a<d$ if $i\leqslant j$ (as in this case $a\in b\oDownset\subseteq d\oDownset$) and $d<c$ otherwise (as $c$ being an element of a forcing path with $c> b\in B_i$ implies $c > B_1\cup\ldots\cup B_{i-1}\supseteq B_j$). This contradiction to ${d}\parallel{a,c}$
shows that \PP is $\ThreePlusOne$-free so it is a semi-order.
\end{proof}
\begin{proof}[Proof of Fact \ref{fact-width-P-is-w}]
To prove that $\fWidth(\PP)=w$ consider any antichain $X$ in $\PP$. We will
show that $\abs{X}\leq w$. Let $m\in X$ be the point with a maximal
down set among points in $X$. We distinguish between three cases:

 If $m\in A$, then $X\subseteq A$ and $\abs{X}\leq\abs{A}=w$.
 If $m\in B_i$, then $X\subseteq B_1\cup\ldots\cup
  B_i\cup(A-B_i\oDownset)$ and
  $\abs{X}\leq(x_0-x_1)+\ldots+(x_{i-1}-x_i)+(w-(x_0-x_i))=w$.

  If $m\in C_0$, then $X\subseteq \bigcup_i B_i$ and $\abs{X}\leq
  \abs{C_0}+\abs{B_1}+\ldots+\abs{B_{k+1}}=(x_0-x_1)+(x_0-x_1)+\ldots+(x_k-x_{k+1})
  =x_0+x_0-x_1\leq w$ by (\ref{equ-Ik}).

  The last and most interesting case is when  $m\in C_j$ for $j>0$. We may write $m\oDownset =
  A\cup(B_1\cup\ldots\cup B_{j-1}\cup D^t_{j})$ where again $D^t_{j}$ is the
  set $D_j$ at the moment when $m$ was inserted. When $m$ is the starting
  point of a forcing path we have $D^t_{j} = \emptyset$.  Clearly, $X\subseteq
  Y\cup(\bigcup_i B_i-m\oDownset)$ where $Y=\set{c\in \bigcup_i C_i:c\oDownset\subseteq m\oDownset}$.

  Since the starting points of forcing paths in $Y$ were introduced in phases $0$ to $j-1$,
  their total number is $(x_0-x_1)+(x_1-x_2)+\ldots+(x_{j-1}-x_{j})=x_0-x_{j}$. The
  introduction of each $c\in Y$ being not a starting point of a forcing
  path is preceded by an extension of some $D_i$ for $1\leq i\leq
  j$. Therefore the number of non-starting points in $Y$ is bounded by
  $\abs{D_1}+\ldots+\abs{D_{j-1}} + \abs{D^t_{j}}$.
  To simplify this expression we first prove the following bound:
  $$
  \abs{D_i}\leqslant x_i.
  $$

For the proof of the inequality note that the set $D_i$ is enlarged
only if a point from a forcing path is assigned to a chain with a top
from $B_i$. This can only happen for forcing paths presented in phases following phase $i$. Each forcing path can contribute at most one point to
$D_i$. There are $(x_i-x_{i+1})+\ldots+(x_{k}-x_{k+1})$ forcing paths 
in phases presented after phase $i$. Since $x_{k+1}=0$ this is not greater than $x_i$, as claimed.
  
  Collecting pieces from above we get $\abs{Y} \leq
  (x_0-x_j)+\abs{D_1}+\ldots+\abs{D_{j-1}}+\abs{D_i^t}
  \leq (x_0-x_j)+ x_1 +\ldots+ x_{j-1} + \abs{D_j^t}$.

   Recall that $\bigcup_i B_i-m\oDownset=(B_j-D_j^t)\cup B_{j+1}\cup\ldots B_{k+1}$. All this finally yields: 
\begin{eqnarray*}
     \abs{X}&\leq& \abs{Y}+\abs{\bigcup\nolimits_i B_i-m\oDownset}\\
     &=& [(x_0-x_j)+ x_1 +\ldots+ x_{j-1} + \abs{D_j^t}] \\
     & & \hbox to 6mm{\hss}+ [(x_{j-1}-x_j-\abs{D_j^t})+
     (x_j-x_{j+1})+ \ldots+ (x_k-x_{k+1})]\\
     &=& x_0 +x_1 +\ldots+ x_{j-1}+ (x_{j-1}-x_j)
\end{eqnarray*}
From (\ref{equ-Ik}) we know that this last expression is not greater than $w$.
\end{proof}
\begin{proof}[Proof of Fact \ref{fact-number-of-chains-forced}]
We will prove that Algorithm is forced to use at
least $w+x_0$ chains on $\PP$. 
First, we show that
$$ \textnormal{all points in $A-B_{k+1}\oDownset$ are tops of the
  chains to which they are assigned.}
$$ 
For the proof of this statement first consider points in \PP dominating
$A-B_{k+1}\oDownset$. These are exactly the points in $\bigcup_i C_i$. Recall that if Algorithm produced a skip chain and assigned $c\in \bigcup C_i$ to a chain whose top was equal to $a\in A$ then $c$ ends a forcing path. If this forcing path was played in phase $j$, then Spoiler later presented $B_{j+1}$ in such a way that $B_{j+1}>a$ and therefore $a\in B_{j+1}\oDownset\subseteq B_{k+1}\oDownset$ so $a\not\in A-B_{k+1}\oDownset$ and we are done.

Consider the set $E$ of end points of
forcing paths presented in the game. The key fact is that
all points in $(A-B_{k+1}\oDownset)\cup \bigcup_i B_i \cup E$ are covered with
distinct chains. Indeed, we have shown that chains in $A-B_{k+1}\oDownset$ 
are tops of the chains. End points of forcing
paths are, by definition, in a chain that is not used in $\bigcup_i B_i$. 
Recall that
\begin{enumeratei}
\item $\abs{A-B_{k+1}\oDownset}=w-x_0$,
\item
  $\abs{B_1}+\ldots+\abs{B_{k+1}}
  =(x_0-x_1)+\ldots+(x_k-x_{k+1}) =x_0$,
\item $\abs{E}=(x_0-x_1)+(x_1-x_2)+\ldots+(x_k-x_{k+1})=x_0$.
\end{enumeratei}
Hence $\abs{A-B_{k+1}\oDownset}+\sum_i\abs{B_i}+\abs{E}=w+x_0$ which gives the
lower bound on the number of chains used by Algorithm.
\end{proof} % of Proposition~\ref{prop:strategy-lower}.
\end{proof}
\begin{proof}[Proof of Proposition~\ref{prop:solution_I_k}.]
We will show that for any $w$ there is a solution of \eqref{equ-Ik} 
with $x_0=\lfloor(\phi-1)\cdot w\rfloor$. Consider the following sequence:
\begin{align*}
x_0&=\lfloor(\varphi-1)\cdot w\rfloor,\\
x_{j+1}&=\lfloor(\varphi-1)\cdot (w-x_0-\ldots-x_j)\rfloor.
\end{align*}
\noindent
Note that for any $0\leqslant a\leqslant x$ we have $\lfloor(\varphi-1)(x-a)\rfloor\leqslant (\varphi-1)x-(\varphi-1)a<(\varphi-1)x$ and thus $\lfloor(\varphi-1)(x-a)\rfloor\leqslant \lfloor(\varphi-1)x\rfloor$.
It implies that the sequence of $x_j$'s is decreasing. Moreover, it eventually gets to zero since the partial sum $x_0+\ldots+x_j$ is getting larger.
%\noindent Note that the sequence of $x_j$'s is strictly decreasing 
%until it gets to be zero. 
In particular there is a $k$ such that
$x_0\geqslant\ldots\geqslant x_k>x_{k+1}=0$. It is easy to verify that the
sequence is a solution of \eqref{equ-Ik}, indeed
\[
x_j\leq (\phi-1)(w-x_0-\ldots-x_{j-1})=\tfrac{\phi}{\phi+1}(w-x_0-\ldots-x_{j-1}).
\]
Multiplying this by $\phi+1$, moving the term $\phi\,x_j$ to the right hand side, adding $x_0+\ldots+x_j$ on both sides
and $-w + w$ on the right side we get
\[
x_0+\ldots+x_{j-1}+2x_j\leq (\phi-1)(w-x_0-\ldots-x_j)+w.
\]
The left side is an integer, therefore we
may take the floor of the right side without affecting the truth.
This results in an inequality from the system \eqref{equ-Ik}:
\[
x_0+\ldots+x_{j-1}+2x_j\leq\lfloor(\phi-1)(w-x_0-\ldots-x_j)\rfloor+w=x_{j+1}+w.
\]

Hence $(x_0,\ldots,x_k)$ is indeed a solution of
\eqref{equ-Ik}, concluding the proof of Proposition~\ref{prop:solution_I_k}.
\end{proof} 

% *************************************************************
\subsection{The Upper Bound}\label{ssec:upper-bound}
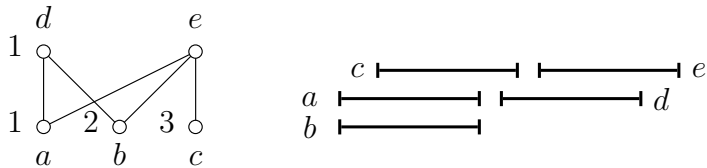
\begin{figure}[htb]
    \begin{center} \ifx\JPicScale\undefined\def\JPicScale{1}\fi
\psset{unit=\JPicScale mm}
\psset{linewidth=0.3,dotsep=1,hatchwidth=0.3,hatchsep=1.5,shadowsize=1,dimen=middle}
\psset{dotsize=0.7 2.5,dotscale=1 1,fillcolor=black}
\psset{arrowsize=1 2,arrowlength=1,arrowinset=0.25,tbarsize=0.7 5,bracketlength=0.15,rbracketlength=0.15}
\begin{pspicture}(0,0)(86.25,18.12)
\rput[b](0,0){$a$}
\rput[b](10,0){$b$}
\rput[b](20,0){$c$}
\rput[b](0,18.12){$d$}
\rput[b](20,18.12){$e$}
\rput[b](-3.75,4.38){1}
\rput[b](6.25,4.38){2}
\rput[b](16.25,4.38){3}
\rput[b](-3.75,14.38){1}
\psline[linewidth=0.15](0,15)(10,5)
\psline[linewidth=0.15](20,15)(10,5)
\rput{0}(10,5){\psellipse[linewidth=0.15,fillcolor=white,fillstyle=solid](0,0)(0.88,0.88)}
\psline[linewidth=0.15](20,15)(20,5)
\rput{0}(20,5){\psellipse[linewidth=0.15,fillcolor=white,fillstyle=solid](0,0)(0.88,0.88)}
\psline[linewidth=0.15](20,15)(0,5)
\rput{0}(20,15){\psellipse[linewidth=0.15,fillcolor=white,fillstyle=solid](0,0)(0.88,0.88)}
\psline[linewidth=0.15](0,15)(0,5)
\rput{0}(0,5){\psellipse[linewidth=0.15,fillcolor=white,fillstyle=solid](0,0)(0.88,0.88)}
\rput{0}(0,15){\psellipse[linewidth=0.15,fillcolor=white,fillstyle=solid](0,0)(0.88,0.88)}
\psline[linewidth=0.4,arrowsize=0 2,arrowlength=0,arrowinset=-0.25,tbarsize=0 5,bracketlength=0,rbracketlength=0]{|-|}(38.75,5)(57.51,5)
\psline[linewidth=0.4,arrowsize=0 2,arrowlength=0,arrowinset=-0.25,tbarsize=0 5,bracketlength=0,rbracketlength=0]{|-|}(38.75,8.75)(57.51,8.75)
\psline[linewidth=0.4,arrowsize=0 2,arrowlength=0,arrowinset=-0.25,tbarsize=0 5,bracketlength=0,rbracketlength=0]{|-|}(43.75,12.5)(62.51,12.5)
\psline[linewidth=0.4,arrowsize=0 2,arrowlength=0,arrowinset=-0.25,tbarsize=0 5,bracketlength=0,rbracketlength=0]{|-|}(65.01,12.5)(83.75,12.5)
\psline[linewidth=0.4,arrowsize=0 2,arrowlength=0,arrowinset=-0.25,tbarsize=0 5,bracketlength=0,rbracketlength=0]{|-|}(60.01,8.75)(78.75,8.75)
\rput(35.01,5){$b$}
\rput(35.01,8.75){$a$}
\rput(41.25,12.5){$c$}
\rput(86.25,12.5){$e$}
\rput(81.25,8.75){$d$}
\end{pspicture} \end{center}
    \caption{Order \PP\ with its unit interval representation and
the chain partition \GAMMA\ of points $a, b, c, d$.} \label{figure:natural}
\end{figure}

\begin{figure}[!ht]
    \begin{center} \ifx\JPicScale\undefined\def\JPicScale{1}\fi
\psset{unit=\JPicScale mm}
\psset{linewidth=0.3,dotsep=1,hatchwidth=0.3,hatchsep=1.5,shadowsize=1,dimen=middle}
\psset{dotsize=0.7 2.5,dotscale=1 1,fillcolor=black}
\psset{arrowsize=1 2,arrowlength=1,arrowinset=0.25,tbarsize=0.7 5,bracketlength=0.15,rbracketlength=0.15}
\begin{pspicture}(0,0)(87.5,19.38)
\psline[linewidth=0.15,linestyle=dashed,dash=1 1](30,15)(20,5)
\psline[linewidth=0.15](20,15)(10,5)
\psline[linewidth=0.15](10,15)(10,5)
\rput[b](0,0){$a$}
\rput[b](10,0){$b$}
\rput[b](20,0){$c$}
\rput[b](0,18.12){$d$}
\rput[b](-3.75,4.38){1}
\rput[b](6.25,4.38){2}
\rput[b](16.25,4.38){3}
\rput[b](-3.75,14.38){1}
\psline[linewidth=0.15](0,15)(10,5)
\rput{0}(10,5){\psellipse[linewidth=0.15,fillcolor=white,fillstyle=solid](0,0)(0.88,0.88)}
\psline[linewidth=0.15](20,15)(20,5)
\rput{0}(20,5){\psellipse[linewidth=0.15,fillcolor=white,fillstyle=solid](0,0)(0.88,0.88)}
\psline[linewidth=0.15](20,15)(0,5)
\psline[linewidth=0.15](0,15)(0,5)
\rput{0}(0,5){\psellipse[linewidth=0.15,fillcolor=white,fillstyle=solid](0,0)(0.88,0.88)}
\rput{0}(0,15){\psellipse[linewidth=0.15,fillcolor=white,fillstyle=solid](0,0)(0.88,0.88)}
\rput{0}(10,15){\psellipse[linewidth=0.15,fillcolor=white,fillstyle=solid](0,0)(0.88,0.88)}
\rput(10,19.38){$q$}
\psline[linewidth=0.4,arrowsize=0 2,arrowlength=0,arrowinset=-0.25,tbarsize=0 5,bracketlength=0,rbracketlength=0]{|-|}(35,5)(53.75,5)
\psline[linewidth=0.4,arrowsize=0 2,arrowlength=0,arrowinset=-0.25,tbarsize=0 5,bracketlength=0,rbracketlength=0]{|-|}(40,8.75)(58.75,8.75)
\psline[linewidth=0.4,arrowsize=0 2,arrowlength=0,arrowinset=-0.25,tbarsize=0 5,bracketlength=0,rbracketlength=0]{|-|}(45,12.5)(63.75,12.5)
\psline[linewidth=0.4,arrowsize=0 2,arrowlength=0,arrowinset=-0.25,tbarsize=0 5,bracketlength=0,rbracketlength=0]{|-|}(66.25,12.5)(85,12.5)
\rput(31.25,5){$b$}
\rput(36.25,8.75){$a$}
\rput(42.5,12.5){$c$}
\rput(87.5,12.5){$e$}
\psline[linewidth=0.4,arrowsize=0 2,arrowlength=0,arrowinset=-0.25,tbarsize=0 5,bracketlength=0,rbracketlength=0]{|-|}(61.25,8.75)(80,8.75)
\rput(82.5,8.75){$d$}
\psline[linewidth=0.4,arrowsize=0 2,arrowlength=0,arrowinset=-0.25,tbarsize=0 5,bracketlength=0,rbracketlength=0]{|-|}(56.25,5)(75,5)
\rput(77.5,5){$q$}
\rput[b](20,18.12){$e$}
\rput{0}(20,15){\psellipse[linewidth=0.15,fillcolor=white,fillstyle=solid](0,0)(0.88,0.88)}
\rput[b](30,18.12){$p$}
\rput{0}(30,15){\psellipse[linewidth=0.15,fillcolor=white,fillstyle=solid](0,0)(0.88,0.88)}
\end{pspicture} \end{center}
    \caption{Point $q$ may be presented by Spoiler in the future,
      point $p$ can not.} \label{figure:natural-2}
\end{figure}
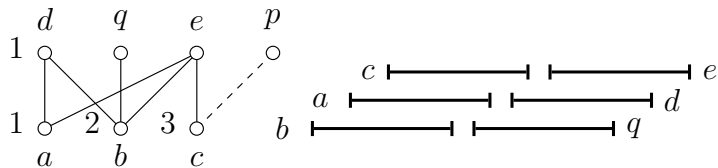

% *************************************************************
Consider a semi-order $\PP=(P, \leqslant)$ with $P = (a,\,
b,\, c,\, d,\, e)$ and the chain partition $\GAMMA : P\setminus\set{e}
\rightarrow \N$ as shown in Fig.~\ref{figure:natural}.
Point $e$ may be covered with a new chain (say, with number $4$) or
with one of the chains already used. In the latter case Algorithm may
choose between $2$ and $3$. We say that chain $\alpha$ is \emph{valid}
for a new point~$x$ extending an already partitioned order \PP\ if $x$
dominates all points from~$\alpha$ in \PP. We claim that among the
valid chains $2$ and $3$ defining $\GAMMA(e) = 3$ is the better
choice. Indeed, any future point $p$ presented by Spoiler and
dominating $c$ will also dominate $b$ (otherwise, \PP\ would have a
\TwoPlusTwo\ configuration which is forbidden in interval orders). On
the other hand, Spoiler may play $q$ greater than $b$ but remaining
incomparable to~$c$ (see Fig.~\ref{figure:natural-2}).
Hence, using the chain of $c$ for $e$ leaves more options for the
future. Whenever the chains of two points $x$ and $y$ are valid and
$x\oUpset\subsetneq y\oUpset$ then it seems safer to use the
chain of $x$. Our Algorithm \ALG will go along this
intuition.

Suppose that Spoiler introduces a semi-order $\PP=(P,\leqslant)$ with
a presentation order $P=(p_1,\ldots,p_n)$. We refer to the chains used by \ALG as \ALG-chains or just chains.
With $\PP_{x}$ we denote the order consisting of all points presented
prior to $x$.  We say that a chain $\alpha$ is {\it valid} for point
$x$ if $\fTop(\alpha)$ in $\PP_{x}$ is below $x$ in \PP. The options
of \ALG are to put $x$ into some valid chain or into a new chain. We are ready
to describe Algorithm's strategy. Let $x$ be a new point presented by
Spoiler.
\smallskip

\FreeItem{\textbf{Algorithm \ALG:}} 
If there is a valid chain for $x$, then put $x$ into 
a valid chain $\alpha$
such that $\fTop(\alpha)\oUpset\subseteq\fTop(\beta)\oUpset$ in $\PP_x$
for all valid chains $\beta$. Otherwise, if there is no valid
chain, use a new chain for $x$. 
\medskip

\noindent
\ALG is a greedy algorithm, i.e., it uses a new chain
only when it is left with no other option. Note also that \ALG has
some freedom in choosing a chain for a new point $x$ as there may be
many tops of valid chains for $x$ with the same minimum up set in
$\PP_x$.

The bound on the performance of \ALG
on width $w$ is stated in the following proposition. This is the upper bound
part for Theorem~\ref{thm-main}.

\begin{proposition}\label{prop:upper}
\ALG uses at most $\phi\cdot w$ chains on any up-growing semi-order of width at most $w$.
\end{proposition}

Suppose Spoiler presents a semi-order $\PP=(P,\leqslant)$ with the presentation
order $P=(p_1,\ldots,p_n)$ and $\fWidth(\PP)\leqslant w$.  We may
assume that \ALG uses a new chain for the last point $p_n$. 
If this were not the case Spoiler could stop the game 
earlier and the number of chains used
by \ALG would remain the same.

We partition \PP into layers. These layers will in some way reflect the preferences of \ALG during the game. The point $p\in\PP$ is \emph{significant} if $p$ dominates at least one maximal point of $\PP_p$. By the linearity of down sets of an interval order $\PP$ this is equivalent to the fact that $p$ has the largest down set at the moment of its presentation. Let $e_1,\ldots,e_{m-1}$ be the significant points of $\PP$, sorted with respect to the presentation order (If $\PP$ has no significant points then $\PP$ is an antichain and the thesis is trivial). These points define the partition of $\PP$ into layers as follows (put $e_0\oDownset=\emptyset$):
\begin{align*}
D_i &= e_{i}\oDownset-e_{i-1}\oDownset,\quad\textnormal{for }1\leq i< m,\\
D_m &=P-e_{m-1}\oDownset.
\end{align*}
\noindent Thus, $D_i$ (for $1\leq i<m$) is exactly the set of maximal points of $\PP_{e_i}$ covered by $e_i$. Here is a list of helpful and easy properties of the $D_i$'s.
\begin{fact}\hfill\label{fact:layers}
\begin{enumeratei}
\item $d_1\oUpset\supsetneq d_2\oUpset\supsetneq\ldots\supsetneq
  d_m\oUpset$, for all $d_i\in D_i$.\label{enu:up-incl}
\item $D_m$ is exactly the set of maximal points of \PP. In
  particular, $p_n\in D_m$.\label{enu:empty upsets of Dm}
\item $D_i$ is an antichain, for every
  $i$.\label{enu:Di-is-an-antichain}
\item If $d_i\in D_i$, $p\in P$ and $d_i<p$ then $D_1\cup\ldots\cup
  D_{i-1}\subseteq
  p\oDownset$.\label{enu:packing-layers-into-down-sets}
\item If $d_i\in D_i$, $p\in P$ and $d_i\nless p$ then
  $p\oDownset\subseteq D_1\cup\ldots\cup
  D_i$.\label{enu:packing-down-sets-into-layers}
\item If $d_i\in D_i$, $p\in P$ and $p$ is presented prior to $d_i$ then $p\oDownset\subseteq D_1\cup\ldots\cup D_{i-1}$.
\label{enu:better-packing-down-sets}
\end{enumeratei}
\end{fact}
\begin{proof}

\itemref{enu:up-incl} 
From $e_{i} \in d_i\oUpset$ but $e_{i} \not\in d_j\oUpset$ for all
$j > i$ and the linear order on the up-sets of elements we obtain
$d_i\oUpset\supsetneq d_j\oUpset$.

\itemref{enu:empty upsets of Dm} 
The last significant point $e_{m-1}$ has the largest down set in \PP. This means that all points outside $e_{m-1}\oDownset$, namely $D_m=P-e_{m-1}\oDownset$, have empty up sets.

\itemref{enu:Di-is-an-antichain}
This follows from \itemref{enu:empty upsets of Dm} and the fact that $D_i$ is the set of maximal points of $\PP_{e_i}$.

\itemref{enu:packing-layers-into-down-sets}
Suppose $i>1$ since for $i=1$ the claim is obvious. Clearly, $d_i\not<e_{i-1}$. But if $d_i > e_{i-1}$ then $e_{i-1}\oDownset\subsetneq d_i\oDownset$. Therefore $d_i$ is the next significant point after $e_{i-1}$ or $e_i$ was presented before $d_i$. Both cases are impossible as $D_i\subseteq e_i\oDownset$, thus $d_i\parallel e_{i-1}$. Now, if $d_i<p$ then by the linearity of down sets we obtain $D_1\cup\ldots\cup D_{i-1}=e_{i-1}\oDownset\subsetneq p\oDownset$.

\itemref{enu:packing-down-sets-into-layers} 
Suppose that $p>d$ for some $d\in D_j$, $j>i$. Then
\itemref{enu:packing-layers-into-down-sets} guarantees that
$D_i\subseteq p\oDownset$. This implies  $d_i < p$, 
contradicting the assumptions.

\itemref{enu:better-packing-down-sets}
If $i=m$ then the thesis is trivial as $D_m$ is the set of maximal points of \PP. Thus, suppose $i<m$. Recall that $e_i$ is the first presented point  which dominates every point from $D_i$. Now as $p$ is presented prior to $d_i$ and $d_i$ precedes $e_j$ for $j\geq i$, we get that $p$ cannot dominate any point from $D_j$.
\end{proof}

With the next fact we prove that \ALG always chooses a chain whose top
is in the highest of the layers which contain tops of valid chains.

\begin{fact}\label{fact:ALG-preferences}
Let $d_i\in D_i$, $d_j\in D_j$, $i<j$ and $d_i,d_j<x$. If $d_i$ is an
\ALG-top in $\PP_x$ and \ALG uses its chain on $x$ then $d_j$ is not
an \ALG-top in $\PP_x$.
\end{fact}
\begin{proof}
At the moment when Spoiler presents $x$, both $d_i$ and $d_j$ are
already introduced (\PP is up-growing). Since $e_{j-1}$ was presented before
$d_j$, $d_i \in e_{j-1}\oDownset$ and $d_j \not< e_{j-1}$
we conclude that $d_i\oUpset\supsetneq d_j\oUpset$ in $\PP_x$. Hence 
if both $d_i$ and $d_j$ are valid tops for $x$ then \ALG has a  preference 
for using the chain of $d_j$ for $x$.
\end{proof}

Before we proceed with the proof we introduce the concept of a
{predecessor} and a {successor} of a point with respect to some fixed
chain partition. Let \CC be the chain partition of
$\PP=(P,\leqslant)$. For $p\in P$ with $p\in C$ for a chain $C\in\CC$
we define:
\begin{enumeratei}
 \item The \emph{predecessor} of $p$ in \CC is the point preceding $p$
   in $C$ (if $p$ is the least point in $C$ then the predecessor of
   $p$ does not exist).
 \item The \emph{successor} of $p$ in \CC is the point succeeding $p$ in
   $C$ (if $p$ is the largest point in $C$ then the successor of $p$
   does not exist).
\end{enumeratei}
We fix an optimal chain partition \OO of \PP. Since
$\fWidth(\PP)\leqslant w$ this partition consists of at most $w$
chains. With respect to this partition we denote the predecessor and
the successor of $p\in P$ by $o^-(p)$ and $o^+(p)$,
respectively. Analogously, we refer to the predecessor and the
successor of $p$ with respect to the the chain partition constructed
by \ALG as $\alg^-(p)$ and $\alg^+(p)$.

We arrive at the key concept of the proof: the
{alternating paths}. Each such path starts at the bottom of an
\ALG-chain. We propose to understand it as a chain of
events originating from the starting bottom. By counting the
number of such paths we will get a bound on the number of chains used
by \ALG.

For each \ALG-chain $\alpha$ define an \emph{alternating path}
$q=(q_0,\ldots)$ as follows:
\begin{enumeratei}
 \item $q_0$ is the bottom point of $\alpha$,
 \item $q_{2i+1}=o^-(q_{2i})$, if $o^-(q_{2i})$ does exist,
 \item $q_{2i+2}=\alg^+(q_{2i+1})$, if $\alg^+(q_{2i+1})$ does exist.
\end{enumeratei}
% \medskip

\noindent
We claim that for each path $q$ all the $q_{2i}$'s are pairwise distinct and so are
all the $q_{2i+1}$'s. Indeed, note that $q_0\neq q_{2i}$ for $i>0$ as
$q_{2i}=\alg^+(q_{2i-1})$ is not a bottom of an \ALG-chain, while
$q_0$ is defined as a bottom. Now suppose that the claim does not hold
and consider the least $i$ such that $q_i=q_j$ for some $j>i$ where
$i$ and $j$ have the same parity. If $i$ and $j$ are even we get
$q_{i-1}=\alg^-(q_i)=\alg^-(q_j)=q_{j-1}$ and if they are odd we get
$q_{i-1}=o^+(q_i)=o^+(q_j)=q_{j-1}$.  In both cases this contradicts
the choice of $i$. This fact implies that the alternating paths are
finite.  Note that an alternating path $q=(q_0,\ldots,q_l)$ is
uniquely determined by any $q_i \in q$ together with the information
whether $q_i$ is an odd or a even element. Altogether we have proven
the following fact:

\begin{fact}\label{fact-disjoint-paths}
For an alternating path $q=(q_0,\ldots,q_l)$ all the $q_i$'s with the
same parity of indices are pairwise distinct, i.e.\ $q_{2i}\neq
q_{2j}$ and $q_{2i+1}\neq q_{2j+1}$ for $i\neq j$. Moreover, each
$p\in P$ occurs in at most two alternating paths: once with an odd
index and once with an even index.
\end{fact}

For an alternating path $q=(q_0,\ldots,q_l)$ we call the $q_{2i}$'s
the \emph{up-points} of $q$ and the $q_{2i+1}$'s the
\emph{down-points} of $q$.  An alternating path $q=(q_0,\ldots,q_l)$
is an \emph{up-path} if its last point is an up-point, otherwise, $q$
is a \emph{down-path}. Note that an up-path connects the bottom of an
\ALG-chain with the bottom of an $\OO$-chain, hence, there are at most
$w$ up-paths (see Fact~\ref{fact-U}). The goal is to bound the number of
down-paths.

From our perspective the important layers in the partition of
$P=D_1\cup\ldots\cup D_m$ will be those containing at least one end
point of a down-path. Define
$$
 I = \set{\; i:\ \textnormal{there is a down-path ending in $D_i$}\;}
   = \set{\; i_0<i_1<\ldots<i_s\;}.
$$
\noindent Note that $m\not\in I$ being an end point of a down-path has a
non-empty up set, while  up sets of all points in $D_m$ are empty (see
Fact \ref{fact:layers}.\itemref{enu:empty upsets of Dm}). This means
\begin{equation}
 i_s<m.\label{equ:is<m}
\end{equation}
\noindent 
From the definition of~$I$ we immediately obtain: 
\begin{fact}\label{fact:ALG-top-in-Dij}
There is an \ALG-top in every $D_{i_j}$, for $0\leq j\leq s$.
\end{fact}

The next fact basically states that
$D_{i_0}\cup\ldots\cup D_m$ induce an order of height at most $3$.
\begin{fact}\label{fact:several-layers-form-an-antichain}
$D_{i_0+1}\cup\ldots\cup D_{m-1}$ is an antichain.
\end{fact}
\begin{proof}
In order to get contradiction suppose that there are $d,\,d'\in
D_{i_0+1}\cup\ldots\cup D_{m-1}$ and that $d<d'$. As $d'$ dominates a
point from a layer higher than $D_{i_0}$, we get from Fact
\ref{fact:layers}\itemref{enu:packing-layers-into-down-sets} that
$D_{i_0}\subseteq d'\oDownset$. On the other hand, since $d'\not\in
D_m$ there must be some point $d''>d'$ (see Fact
\ref{fact:layers}\itemref{enu:empty upsets of Dm}). Fix some $t\in D_{i_0}$
that remains an \ALG-top throughout the game (such $t$ exists by Fact
\ref{fact:ALG-top-in-Dij}). Recall that $p_n$, the last point
presented by Spoiler, is incomparable with $t$ as otherwise $p_n$
would not be assigned to a new \ALG-chain. Together this shows that $p_n$
and  $t<d'<d''$ form a \ThreePlusOne-configuration.
This is impossible since $\PP$ is a semi-order.
\end{proof}

We now introduce variables that count the number of paths with respect
to their end points. Define
\begin{tabbing}
\quad$x_U$\quad\= \kill
\quad$x_U$\>the total number of up-paths,\\
\quad$x_j$\>the number of down-paths ending in 
            $\bigcup_{i\geqslant i_j}D_i$, for $j=0,\ldots,s$,\\
\quad$x_{s+1}$\>$=0$.
\end{tabbing}
\smallskip

\noindent
In terms of these variables the number of chains used by \ALG 
can be expressed as 
\[
x_U+x_0.
\]
\begin{fact}\label{fact-U}
$x_U\leq w$.
\end{fact}
\begin{proof}
Consider the set $U$ of end points of up-paths. By Fact
\ref{fact-disjoint-paths} these end points are pairwise distinct and
$\abs{U}=x_U$. As each point in $U$ is a bottom point of its chain in
$\OO$ these points belong to different chains in \OO. Therefore,
$x_U=\abs{U}\leq\abs{\OO}\leq w$.
\end{proof}

Throughout the rest of the paper our efforts are targeted on bounding
the number~$x_0$ of down-paths.  In the following two lemmas we will present a system of inequalities involving $x_0$ and the other~$x_i$'s. From these
inequalities we will derive the desired constraints on the value of $x_0$.
\begin{lemma}\label{lemma-0}
$x_0+x_0-x_1\leqslant w$.
\end{lemma}
\begin{lemma}\label{lemma-1}
$x_0+x_1+\ldots+x_j+(x_j-x_{j+1})\leqslant w$, for $j=1,\ldots,s$.
\end{lemma}

We need some preparations for the proofs of the lemmas. In the
counting arguments on which the proofs are based we will be often showing that certain sets are disjoint. Below we introduce a criterion that
will help us do the bookkeeping.

Point $p\in P$ is a \emph{good} point if $o^-(p)$ exists and
$o^-(p)$ is an \ALG-top at the moment when $p$ is presented, i.e.,
$o^-(p)$ is a top of its \ALG-chain in $\PP_p$. Simple enough, point $p$ is considered \emph{bad} if it is not good. 

Fact \ref{fact-last-but-one-point-is-good}
states that the penultimate point of a down-path is always good. Fact
\ref{fact:bunch-of-not-good-up-points} says that if a down-path ends
in $D_{i_j}$ then it contains  $j+1$ bad up-points. 
Fact \ref{fact-injection} is a technical statement used in the proof of 
Lemma~\ref{lemma-1}.

\begin{fact}\label{fact-last-but-one-point-is-good}
The penultimate point of a down-path is a good point and lies in $D_m$.
\end{fact}
\begin{proof}
Let $q=(q_0,\ldots,q_{l-1},q_l)$ be a down-path. We have
$q_l=o^-(q_{l-1})$ and since $q$ ends with $q_l=o^-(q_{l-1})$ 
this point must be an \ALG-top in \PP. This shows that $q_{l-1}$ is good.

To prove that $q_{l-1}\in D_m$ we are going to prove an equivalent
condition that the up set of $q_{l-1}$ is empty (see Fact
\ref{fact:layers}\itemref{enu:empty upsets of Dm}). Recall that the
last point presented in \PP, namely $p_n$, receives a new
$\ALG$-chain. This means that there is no valid chain for $p_n$ and in
particular, $p_n\ngtr q_l$. From $p_n\oUpset=\emptyset$ we get
$p_n\parallel q_l$. Now, if there was a point $x>q_{l-1}$ then
$p_n,q_l,q_{l-1},x$ would form a \ThreePlusOne-configuration. 
This is impossible since $\PP$ is a semi-order.
\end{proof}

\begin{fact}\label{fact:bunch-of-not-good-up-points}
A down-path $q$ ending in $D_{i_j}$ contains $j+1$ bad up-points
$\set{y_0,\ldots,y_j}$ such that 
$o^-(y_k)\in D_{i_k}$, for $0\leq k\leq j$.
\end{fact}
\begin{proof}
Fix $k$ and define $y_k$ to be the first up-point in $q$ such that
$o^-(y_k)\in D_l$ for some $l\geq i_k$. Such point does exist, as the
penultimate point in $q$ is a candidate for $y_k$.
\smallskip

\noindent{\it Claim.} $y_k$ is a bad point.
\smallskip

Suppose $y_k$ is the first point of $q$.
In this case $\ALG$ uses a new chain on $y_k$. Since \ALG is
greedy there is no valid chain for $y_k$ at the moment it is
presented. In particular, $o^-(y_k)$ is not an \ALG-top in $\PP_{y_k}$
and therefore $y_k$ is a bad point. 

If $y_k$ is not the first point of
$q=(\ldots,p,o^-(p),y_k,o^-(y_k),\ldots)$, then we know that
$p$ did not qualify for $y_k$ and therefore $\alg^-(y_k)=o^-(p)\in
D_1\cup\ldots\cup D_{i_k-1}$. From Fact \ref{fact:ALG-preferences}
it follows that $o^-(y_k)$ is not an \ALG-top in $\PP_{i_k}$ and again
$y_k$ is bad. This proves the claim.
\smallskip

It remains to show that $o^-(y_k)\in D_{i_k}$.
Fix an \ALG-top $t\in D_{i_k}$
(Fact \ref{fact:ALG-top-in-Dij}). Note that $t\nless y_k$ as otherwise
\ALG would have given preference to the chain of $t$ instead of the one it
used for $y_k$. From Fact
\ref{fact:layers}\itemref{enu:packing-down-sets-into-layers} we get
$y_k\oDownset \subseteq D_1\cup\ldots\cup D_{i_k}$. From this and the
definition of $y_k$ we conclude $o^-(y_k)\in D_{i_k}$.
\end{proof}

\begin{fact}\label{fact-injection}
For every $j$ $(0\leq j\leq s)$ and down-path $q$ there is an up-point
$u\in q$ such that one of the following two conditions is true:
\begin{enumeratei}
\item $u\in D_{i_j+1}\cup\ldots\cup D_m$ and $o^-(u)\in D_{i_0}$, or\label{enu:V}
\item $u$ is good and $o^-(u)\in D_{i_0+1}\cup\ldots\cup D_{i_j-1}$.\label{enu:W}
\end{enumeratei}
\end{fact}
\begin{proof}
Fix $j$ and a down-path $q$. Let $a$ be the last up-point in $q$
such that $o^-(a)\in D_k$ for some
$k\leq i_0$. There is such a point as by Fact
\ref{fact:bunch-of-not-good-up-points} each down-path has an up-point
$y$ with $o^-(y)\in D_{i_0}$.

First we are going to prove that $o^-(a)\in D_{i_0}$. It is trivial if
$o^-(a)$ is the last point of $q$ as the first layer with an end of a
down-path is $D_{i_0}$. If $o^-(a)$ is not the last point then let
$b$ be the next point of $q$,
i.e.\ $q=(\ldots,a,o^-(a),b,o^-(b),\ldots)$. From the way point $a$ is
chosen we have $o^-(b)\in D_l$, for some $l>i_0$. By Fact
\ref{fact:layers}\itemref{enu:packing-layers-into-down-sets}
$b>o^-(b)\in D_l$ implies $D_{i_0}\subseteq b\oDownset$. Now, recall
that there is an $\ALG$-top $t\in D_{i_0}$ (Fact
\ref{fact:ALG-top-in-Dij}). In particular, at the moment when $b$ was
introduced the chain of $t\in D_{i_0}$ was valid for $b$. Since \ALG
has chosen $\alg^-(b)=o^-(a)\in D_k$ we get $k\geq i_0$ (Fact
\ref{fact:ALG-preferences}). Since $a$ was chosen such that 
$o^-(a)\in D_k$ for some $k\leq i_0$ we get $k=i_0$.

If $a\in D_{i_j+1}\cup\ldots\cup D_m$ then $a$ fulfills the
condition for the $u$ of \itemref{enu:V} and we are done.
From now on we deal with the case $a\in D_k$ for some $k \leq i_j$.
As $a>o^-(a)\in D_{i_0}$ point~$a$ must be
somewhere in $D_{i_0+1}\cup\ldots\cup D_m$. 
We therefore know that
\begin{equation}
a\in D_{i_0+1}\cup\ldots\cup D_{i_j}.\label{equ:a-in-some-layers}
\end{equation}

\noindent 
Since $i_j<m$ (see \eqref{equ:is<m}) we have $a\not\in D_m$, hence, there is a
point $a'\in P$ with $a'>a$.  Note that $o^-(a)$ can't be the last
point of $q$.  Otherwise, $a$ would be the penultimate point of $q$
which is in $D_m$ (Fact \ref{fact-last-but-one-point-is-good}).  Let
$b$ be the successor of~$o^-(a)$ in $q$,
i.e.\ $q=(\ldots,a,o^-(a),b,o^-(b),\ldots)$. We claim that Spoiler
presents $b$ prior to $a$ in \PP, i.e.\ $b\in \PP_a$.  From the
definition of $a$ it follows that $b>o^-(b)\in D_l$, for some $l>i_0$
and hence $D_{i_0}\subseteq b\oDownset$.  (Fact
\ref{fact:layers}\itemref{enu:packing-down-sets-into-layers}).  At the
moment when $b$ is presented there are at least two valid chains for
$b$, the one actually used by \ALG with its top in $\alg^-(b)=o^-(a)$
and some chain with top $t\in D_{i_0}$ (Fact
\ref{fact:ALG-top-in-Dij}). Recall that the last point $p_n$ presented
in \PP is put into a new \ALG-chain and therefore $p_n\parallel
t$. Now, if $a>t$ then $p_n$ together with $t<a<a'$ would form a
$\ThreePlusOne$-configuration.  Therefore $a\parallel t$ while
obviously $a>o^-(a)$.  Suppose that $a$ is presented by Spoiler prior
to $b$. This implies that at the moment when $b$ is introduced
(i.e.\ in $\PP_b$) we have $t\oUpset\subsetneq o^-(a)\oUpset$ and
therefore \ALG would prefer the chain of $t$ over the chain of
$\alg^-(b)=o^-(a)$ to be used for $b$.  With this contradiction we
have proved the claim that the order of presentation is
$P=(\ldots,b,\ldots,a,\ldots)$.

Let $c$ be the last up-point in $q$ presented by Spoiler prior to $a$
(i.e.\ in $\PP_a$). There is such a point as $b$ is an up-point of $q$
and it is presented prior to $a$. Since $b$ comes after $a$ on $q$
this also holds for $c$, i.e.,
\begin{equation}
q=(\ldots,a,o^-(a),\ldots,c,o^-(c),\ldots).\label{equ:a-after-c}
\end{equation}

\noindent The last step in the proof is to show that $c$ fulfills the
condition for the $u$ of \itemref{enu:W}.

First, we show that $c$ is good. If not, $z=\alg^+(o^-(c))$ would
be defined and had to be presented prior to $c$, so before $a$ as well. But then $z$ would be also an up-point of $q$ which contradicts the choice of $c$.

It remains to prove that $o^-(c)\in D_{i_0+1}\cup\ldots\cup
D_{i_j-1}$. Recall that $a$ is the last up-point of $q$ with
$o^-(a)\in D_k$ for $k\leq i_0$. Since $c$ comes later than $a$ on $q$ 
(see \eqref{equ:a-after-c}) we have $o^-(c)\in D_k$ for some $k\geq
i_0+1$. To bound $k$ from above recall that $a\in D_{i_0+1}\cup\ldots\cup D_{i_j}$ and $o^-(c)$ is presented prior to $a$. Applying this to Fact \ref{fact:layers}.\itemref{enu:better-packing-down-sets} we get that $o^-(c)\in\bigcup_{i<i_j} D_i$. This finishes the proof of Fact~\ref{fact-injection}.
\end{proof}

With these  preparations we are ready for the proofs
of Lemmas \ref{lemma-0} and \ref{lemma-1}.
\medskip

% \noindent{\sl Proof of Lemma~\ref{lemma-0}.}
\begin{proof}[Proof of Lemma~\ref{lemma-0}.]
We are going to construct an antichain in \PP of size
$x_0+x_0-x_1$. This implies the statement.

By Fact \ref{fact:bunch-of-not-good-up-points} a down-path $q$ contains a bad
up-point $y_q$ with $o^-(y_q)\in D_{i_0}$. Collect these points in a set
$Y=\set{y_q:\textnormal{ $q$ is a down-path}}$. 
Since $x_0$ is just the number of down-paths and $Y$ contains a point
from each down-path we get $\abs{Y}=x_0$.
Let $Z$ be the set of penultimate points of down-paths ending
in $D_{i_0}$. The number of such paths is $x_0-x_1$.
hence  $\abs{Z}=x_0-x_1$. From Fact
\ref{fact-last-but-one-point-is-good} we know that points in $Z$ are
good. Summarizing:
\begin{enumeratei}
\item $\abs{Y}=x_0$, all $y\in Y$ are bad and satisfy $o^-(y)\in D_{i_0}$.
\item $\abs{Z}=x_0-x_1$, all $z\in Z$ are good and satisfy $o^-(z)\in D_{i_0}$.
\end{enumeratei}
This implies that $Y$ and $Z$ are disjoint and
\[
x_0+x_0-x_1=\abs{Y}+\abs{Z}=\abs{Y\cup Z}=\abs{o^-(Y\cup Z)}\leq\abs{D_{i_0}}\leq w,
\]
where the last inequality holds as $D_{i_0}$ is an antichain in \PP
(Fact \ref{fact:layers}\itemref{enu:Di-is-an-antichain}).
\end{proof}
% \medskip

% \noindent{\sl Proof of Lemma~\ref{lemma-1}.}  
\begin{proof}[Proof of Lemma~\ref{lemma-1}.]
The basic idea of the proof is similar to the proof Lemma \ref{lemma-0}
but the details are more involved. For fixed $j$ we
construct a set consisting of $x_0+x_1+\ldots+x_j+x_j-x_{j+1}$ points 
such that the points of the set belong to different chains in
\OO. As \OO contains at most $w$ chains, this implies the
inequality with index $j$.

Fix $j$. First we construct a set of size
$x_1+\ldots+x_j$. For a down-path $q$ ending in $D_{i_k}$ put
$r(q)=\min(k,j)$. By Fact \ref{fact:bunch-of-not-good-up-points} each
$q$ contains a set $Y_q$ of $r(q)$ bad up-points $y$ with
$o^-(y)\in D_{i_1}\cup\ldots\cup D_{i_j}$. Let $Y$ be the union of all these sets $Y_q$.
We claim that that all the $Y_q$'s are pairwise disjoint.
This is true because no point occurs in more
than one alternating path as an up-point (Fact
\ref{fact-disjoint-paths}). The claim implies $\abs{Y}=\sum_q\abs{Y_q}$. We
determine the size of $Y$ as follows
\begin{align*}
\abs{Y}&=\sum_{k=1}^s\sum_{\genfrac{}{}{0pt}{2}{q\text{ ending}}{\text{in }D_{i_k}}}\abs{Y_q}
=\sum_{k=1}^{s}\sum_{\genfrac{}{}{0pt}{2}{q\text{ ending}}{\text{in }D_{i_k}}}r(q)
=\sum_{k=1}^{s}\min(k,j)\cdot(x_k-x_{k+1})\\
&= \sum_{k=1}^{j} k\cdot(x_k-x_{k+1}) + \sum_{k=j+1}^{s}j\cdot(x_k-x_{k+1})  
 = x_1+\ldots+x_j - j\cdot x_{s+1}\\
&= x_1+\ldots+x_j.
\end{align*}
For further reference we collect the important properties of $Y$:
\begin{enumeratei}
\item $\abs{Y}=x_1+\ldots+x_j$.
\item All $y\in Y$ are bad and $o^-(y)\in D_{i_1}\cup\ldots\cup D_{i_j} \subseteq D_{i_0+1}\cup\ldots\cup D_{i_j}$.
\end{enumeratei}
% \medskip

\noindent
The second set to consider is:
\[
Z=\set{z:\ \text{there is a down-path } q=(\ldots,z,o^-(z))
           \text{ and $o^-(z)\in D_{i_j}$}}. 
\]
This is the set of the penultimate points of down-paths
ending in $D_{i_j}$. The penultimate points of down-paths are
up-points and hence all distinct (Fact
\ref{fact-disjoint-paths}). From Fact \ref{fact-last-but-one-point-is-good}
we know that all points in $Z$ are good. Summarizing:
\begin{enumeratei}
\item $\abs{Z}=x_j-x_{j+1}$,
\item All $z\in Z$ are good and $o^-(z)\in D_{i_j}$
\end{enumeratei}
% \medskip

\noindent
With a help of Fact \ref{fact-injection} we construct a third
set $U$. Each down-path $q$ contains an up-point $u_q$ satisfying property
\itemref{enu:V} or \itemref{enu:W} of Fact \ref{fact-injection}.
The set $U$ is the collection of all these points.
Since no point is an up-point of more than one path all the
$u_q$'s are distinct. Since there are $x_0$ down-paths we have 
$\abs{U} = x_0$. We partition the set $U = \set{u_q:\text{ $q$ down-path}}$ 
into three parts $U_1$, $U_2$ and $U_3$ as follows:
\begin{enumeratei}
\item $U_1$ is the set of $u\in U$ with $u\in D_{i_j+1}\cup\ldots\cup
  D_{m-1}$ and $o^-(u)\in D_{i_0}$.\label{enu:U1}
\item $U_2$ is the set of $u\in U$ with $u\in D_{m}$ and $o^-(u)\in
  D_{i_0}$.\label{enu:U2}
\item $U_3$ is the set of $u\in U$ such that $u$ is good and
  $o^-(u)\in D_{i_0+1}\cup\ldots\cup D_{i_j-1}$.\label{enu:U3}
\end{enumeratei}

The following properties of sets $Y$, $Z$ and
$U_1,U_2,U_3$ are crucial:
\begin{enumeratenum}
\item The sum of sizes of $U_1$, $U_2$, $U_3$, $Y$ and $Z$ is
  $x_0+x_1+\ldots+x_j+(x_j-x_{j+1})$.
\item $U_3$, $Y$ and $Z$ are disjoint. Indeed points in $Y$ are bad while 
  points in $Z$ and $U_3$ are good. The predecessors in \OO-chains
  show that $Z$ and $U_3$ are disjoint.
\item $o^-(U_3\cup Y\cup Z)\subseteq D_{i_0+1}\cup\ldots\cup D_{i_j}$
  and $U_1\subseteq D_{i_j+1}\cup\ldots\cup D_{m-1}$.
\item Points in $U_2\cup D_{i_0+1}\cup\ldots\cup D_{m-1}$ lie in
  different $\OO$-chains. Indeed, $D_{i_0+1}\cup\ldots\cup D_{m-1}$ is
  an antichain (Fact \ref{fact:several-layers-form-an-antichain})
  and \OO-chains of points from $U_2$ skip all layers between 
  $D_{i_0}$ and $D_m$, i.e., they avoid $D_{i_0+1}\cup\ldots\cup D_{m-1}$.
\end{enumeratenum}
% \medskip

\noindent
We are now ready to complete the proof:
\begin{align*}
x_0+(x_1+\ldots+x_j)+(x_j&-x_{j+1})=\\
^{(1)}&=(\abs{U_1}+\abs{U_2}+\abs{U_3})+\abs{Y}+\abs{Z}\\
^{(2)}&=\abs{U_2}+\abs{U_3\cup Y\cup Z}+\abs{U_1}\\
&=\abs{U_2}+\abs{o^-(U_3\cup Y\cup Z)}+\abs{U_1}\\
^{(3)}&\leq\abs{U_2}+\abs{D_{i_0+1}\cup\ldots\cup D_{m-1}}\\
^{(4)}&\leq\abs{\OO}\\
&\leq w.
\end{align*}
\end{proof} % of Lemma~\ref{lemma-1}.

\noindent
Lemmas~\ref{lemma-0} and \ref{lemma-1} provide us with a system of
inequalities involving all the $x_i$'s:
\[
 x_0+x_1+\ldots+x_j+x_j-x_{j+1}\leqslant w,\ j=0,\ldots,k.
\]
Note that these are the inequalities used for the lower bound.
Here, for the completion of the proof of Proposition~\ref{prop:upper}
we need a final lemma to bound the number $x_0 + w$ of chains
used by Algorithm.

\begin{lemma}
$x_0\leqslant(\phi-1)\cdot w$.\label{cor-x0-bound}
\end{lemma}
\begin{proof}

\noindent 
We weight the $j$th inequality with the Fibonacci number $F_{2(k-j)+1}$
and take the sum of weighted inequalities:
\begin{equation}\label{equ-SUM-2}
\sum_{j=0}^k \sum_{i=0}^j x_i F_{2(k-j)+1} + \sum_{j=0}^k (x_j- x_{j+1}) F_{2(k-j)+1}
    \leqslant w\sum_{j=0}^k F_{2(k-j)+1}.
\end{equation}

\noindent 
Using the well-known Fibonacci identity
$\sum_{j=0}^k F_{2(k-j)+1} = \sum_{j=0}^k F_{2j+1}=
F_{2k+2}$ we can simplify the double-sum:
\[
\sum_{j=0}^k \sum_{i=0}^j x_i F_{2(k-j)+1} = \sum_{i=0}^k x_i
\sum_{j=i}^k F_{2(k-j)+1} = \sum_{j=0}^k x_j F_{2(k-j)+2}.
\]
\noindent 
This and again using the Fibonacci identity on the right hand side
allows us to rewrite inequality \eqref{equ-SUM-2}:
\[
\sum_{j=0}^k x_j F_{2(k-j)+2} + \sum_{j=0}^k x_j F_{2(k-j)+1} -
\sum_{j=1}^{k+1} x_j F_{2(k-j)+3} \leqslant w F_{2k+2}.
\]
Using the Fibonacci recursion and the fact that $x_{k+1}=0$ 
this reduces to
\[
x_0 F_{2k+2} + x_0 F_{2k+1}\leqslant w F_{2k+2}.
\]
This in turn can be rewritten into
\[
x_0\leqslant\frac{F_{2k+2}}{F_{2k+3}}\cdot w\leqslant(\phi-1)\cdot w.
\]
\noindent The last inequality is due to the fact that the sequence
$(\frac{F_{2k+2}}{F_{2k+3}})_{k \geqslant 0}$ is monotonically increasing
with the limit $\phi-1$.
\end{proof}

The statement of Lemma~\ref{cor-x0-bound} was the last piece of the puzzle. The number of chains used by \ALG is bounded by
\[
x_U+x_0\leq w+(\phi-1)\cdot w.
\]
This completes the proof of Proposition~\ref{prop:upper} and hence of Theorem~\ref{thm-main}.

\bibliographystyle{plain} \bibliography{semi}

\begin{thebibliography}{1}

\bibitem{BBM07}
Patrick Baier, Bart{\l}omiej Bosek, and Piotr Micek.
\newblock On-line chain partitioning of up-growing interval orders.
\newblock {\em Order}, 24(1):1--13, 2007.

\bibitem{FKMM-survey}
Bart{\l}omiej Bosek, Stefan Felsner, Kamil Kloch, Tomasz Krawczyk, Grzegorz
  Matecki, and Piotr Micek.
\newblock On-line chain partitions of orders: a survey.
\newblock to appear in ORDER.

\bibitem{Fel97}
Stefan Felsner.
\newblock On-line chain partitions of orders.
\newblock {\em Theoret. Comput. Sci.}, 175(2):283--292, 1997.
\newblock Orders, algorithms and applications (Lyon, 1994).

\bibitem{Kier81}
Henry~A. Kierstead.
\newblock An effective version of {D}ilworth's theorem.
\newblock {\em Trans. Amer. Math. Soc.}, 268(1):63--77, 1981.

\bibitem{Kier86}
Henry~A. Kierstead.
\newblock Recursive ordered sets.
\newblock In {\em Combinatorics and ordered sets (Arcata, Calif., 1985)},
  volume~57 of {\em Contemp. Math.}, pages 75--102. Amer. Math. Soc.,
  Providence, RI, 1986.

\bibitem{KierTro81}
Henry~A. Kierstead and William~T. Trotter, Jr.
\newblock An extremal problem in recursive combinatorics.
\newblock In {\em Proceedings of the Twelfth Southeastern Conference on
  Combinatorics, Graph Theory and Computing, Vol. II (Baton Rouge, La., 1981)},
  volume~33, pages 143--153, 1981.

\bibitem{Moe89}
Rolf~H. M{\"o}hring.
\newblock Computationally tractable classes of ordered sets.
\newblock In {\em Algorithms and Order}, pages 105--193. Kluwer Acad. Publ.,
  Dordrecht, 1989.

\end{thebibliography}

% *************************************************************
%&%&%&%&%&%&%&%&%&%&%&%&%&%&%&%&%&%&%&%&%&%&%&%&%&%&%&%&%&%&%&%&%&%&%&%&%&%&%&%

\end{document}